\documentclass[%
 reprint,superscriptaddress,
amsmath,amssymb]{revtex4-1}
\usepackage{natbib}
\usepackage{graphicx}
\usepackage{dcolumn}
\usepackage{bm}
\usepackage{color}
\usepackage{enumerate}
\usepackage{xcolor}
\usepackage[normalem]{ulem}
\usepackage{braket}
\usepackage{enumitem}
\usepackage[normalem]{ulem}
\usepackage{lipsum}
\usepackage{xfrac}
\usepackage{comment}
\usepackage{makecell}
\usepackage{lipsum}
\usepackage{tabularx}  
\usepackage{amssymb}   
\usepackage{textcomp}  
\definecolor{wine}{RGB}{128, 0, 32}

\newcommand{\expval}[1]{\left\langle #1 \right\rangle}

\usepackage[colorlinks=true,citecolor=blue,linkcolor=blue]{hyperref}

\begin{document}
\renewcommand{\thefigure}{\arabic{figure}}

\title{Vibrational Origin of the Barocaloric Effect in the Spin-Crossover Complex Fe(pap-5NO$_2$)$_2$: A Combined DFT and Mean-Field Study}


\author{Alan de Souza}
\email{almeida.alan@posgraduacao.uerj.br}
\affiliation{Instituto de Física Armando Dias Tavares, Universidade do Estado do Rio de Janeiro, R. São Francisco Xavier 524, Rio de Janeiro, RJ, 20550-013, Brazil}
\affiliation{Neutron Scattering Division, Oak Ridge National Laboratory, Oak Ridge, TN 37831-6475 United States}
\author{Antonio M. dos Santos}
\email{dossantosam@ornl.gov}
\author{Yongqiang Cheng}
\email{chengy@ornl.gov}
\affiliation{Neutron Scattering Division, Oak Ridge National Laboratory, Oak Ridge, TN 37831-6475 United States}

\author{Vinícius S. R. de Sousa}
\email{vinicius@fis.uerj.br}
\affiliation{Instituto de Física Armando Dias Tavares, Universidade do Estado do Rio de Janeiro, R. São Francisco Xavier 524, Rio de Janeiro, RJ, 20550-013, Brazil}

\author{Mario Reis}
\email{marioreis@id.uff.br}
\affiliation{Instituto de Física, Universidade Federal Fluminense, Av. Gal. Milton Tavares de Souza s/n, 24210-346, Niteroi-RJ, Brazil}

\date{\today} 

\begin{abstract}
\textbf{Abstract:}
Despite the barocaloric performance recently reported for the spin-crossover complex Fe(pap-5NO$_2$)$_2$, the microscopic origin of its pressure-induced entropy change remains poorly understood. In particular, the relative contributions of vibrational, configurational, and spin degrees of freedom to the barocaloric response have not yet been quantitatively established. Here, we develop a theoretical framework that combines density functional theory (DFT) calculations with an effective mean field Hamiltonian to investigate the barocaloric effect in this spin-crossover complex. Vibrational frequencies of the low-spin (S = 0) and high-spin (S = 2) states obtained from first-principles calculations are incorporated into a thermodynamic model that explicitly accounts for configurational, spin, and lattice entropy contributions. By establishing a correspondence between the vibrational spectra of both spin states through displacement-vector overlap analysis, we identify low and mid frequency metal–ligand vibrations as the primary microscopic origin of the vibrational entropy change. The proposed model quantitatively reproduces the experimentally reported barocaloric entropy change for a pressure variation of 2 kbar, corresponding to a maximum reversible entropy change of approximately 70 J kg$^{-1}$ K$^{-1}$, and predicts a spin-crossover temperature of $T_{1/2}=313$ K under ambient pressure, in excellent agreement with the experimental value of 308 K. Analysis of the entropy components reveals that lattice entropy associated with molecular vibrations accounts for approximately 84\% of the total barocaloric response, whereas configurational and spin contributions are comparatively small.

\end{abstract}

\maketitle
\section{Introduction}
Caloric effects are based on the isothermal entropy change ($\Delta S_{\mathrm{T}}$) and adiabatic temperature change ($\Delta T_{\mathrm{ad}}$) induced by variations of an external field, such as an electric field (electrocaloric effect) \cite{Torello2022-ht}, a magnetic field (magnetocaloric effect) \cite{De_Oliveira2010-bo}, or pressure (barocaloric effect) \cite{Sun2025-un,De_Oliveira2014-ww,Guan2026-re}. Among these, the barocaloric effect has recently attracted considerable attention due to its intrinsic advantages, as pressure can generally be generated and controlled more easily than strong magnetic or electric fields \cite{Manosa2020-ob}. Furthermore, pressure-induced thermal responses in solids can be remarkably large, particularly when they are associated with first-order phase transitions \cite{Ribeiro2023-ur}. Among the different classes of materials exhibiting such transitions, spin-crossover (SCO) compounds have emerged as particularly attractive candidates due to the strong coupling between their structural and vibrational degrees of freedom. 

In this context, materials exhibiting spin-crossover (SCO) transitions have been proposed as possible candidates for achieving large barocaloric effects \cite{Sandeman2016-kg,Reis2020-xn}. Among them, the most attractive systems are Fe(II)-based octahedral complexes, in which the spin crossover typically occurs from a diamagnetic low-spin (LS) state (S = 0) to a paramagnetic high-spin (HS) state (S = 2). This transition can be induced by several external stimuli, including temperature, pressure, and light irradiation \cite{Gutlich2012-tw,Real2005-lb}. In this regard, theoretical studies have shown that the entropy change associated with the spin-crossover transition arises predominantly from the increased number of accessible vibrational modes in the high-spin state, with the accompanying volume change providing an additional contribution \cite{Ribeiro2019-co,Von_Ranke2021-kq,Von_Ranke2020-sj}. Several Fe(II)-based SCO complexes have therefore been investigated as potential barocaloric materials. Among them, Fe(pap-5NO$_2$)$_2$ has attracted particular attention owing to its spin transition near room temperature and the large structural changes accompanying the SCO process.

In 2015, Olga Iasco \textit{et al.} \cite{Iasco2015-ov} investigated the magnetic and structural properties of Fe(pap-5NO$_2$)$_2$, revealing the presence of a spin-crossover transition (from S = 0 to S = 2) at the Fe(II) center occurring at approximately 300 K under ambient pressure. This transition is accompanied by a significant volume change between two isostructural monoclinic phases, while preserving the overall crystal symmetry. Such a large volume discontinuity is particularly relevant from a barocaloric perspective, since pressure-induced entropy changes are strongly enhanced in materials undergoing volume-driven first-order transitions. Consequently, Fe(pap-5NO$_2$)$_2$ emerged as a viable candidate for barocaloric applications, motivating further investigations under hydrostatic pressure. 

In 2025, David Gracia \textit{et al.} \cite{Gracia2025-pj} investigated the pressure dependence of the spin-crossover transition in this compound, as well as its barocaloric response, by combining X-ray diffraction and high-pressure calorimetric measurements. Their results showed that the spin-crossover temperature shifts linearly toward higher temperatures with increasing pressure, while the thermal hysteresis widens simultaneously. Since large hysteresis can hinder the reversibility required for practical refrigeration applications, the authors evaluated the conditions under which the barocaloric cycle could operate reversibly. They determined that reversible operation could be achieved within a pressure window of approximately 1.2 kbar. Using this criterion, they calculated the reversible barocaloric figures of merit for a pressure change of $\Delta p = 2$ kbar. Under these conditions, the maximum adiabatic temperature change ($\Delta T$) decreased from 20 K to 14 K, whereas the maximum entropy change ($\Delta S$) exhibited only a modest reduction, decreasing from 79 to 70 J.kg$^{-1}$.K$^{-1}$.

Despite the remarkable experimental barocaloric performance reported for Fe(pap-5NO$_2$)$_2$, a detailed theoretical understanding of the microscopic mechanisms governing its pressure induced entropy changes remains limited. In particular, the relative contributions of vibrational, magnetic, and structural degrees of freedom to the barocaloric response have not yet been fully clarified. Therefore, a comprehensive theoretical framework capable of describing the barocaloric properties of this compound is still lacking. In this work, we present a theoretical description of the barocaloric effect in Fe(pap-5NO$_2$)$_2$ based on the thermodynamic properties associated with its spin crossover transition. By combining effective Hamiltonian in the mean field approximation and density functional theory, we evaluate the entropy variation induced by pressure and analyze the microscopic origin of the observed barocaloric response.

\section{Material description}
\label{Section:material_description}

\begin{figure}[!htb]
    \centering
    \includegraphics[width=1\linewidth]{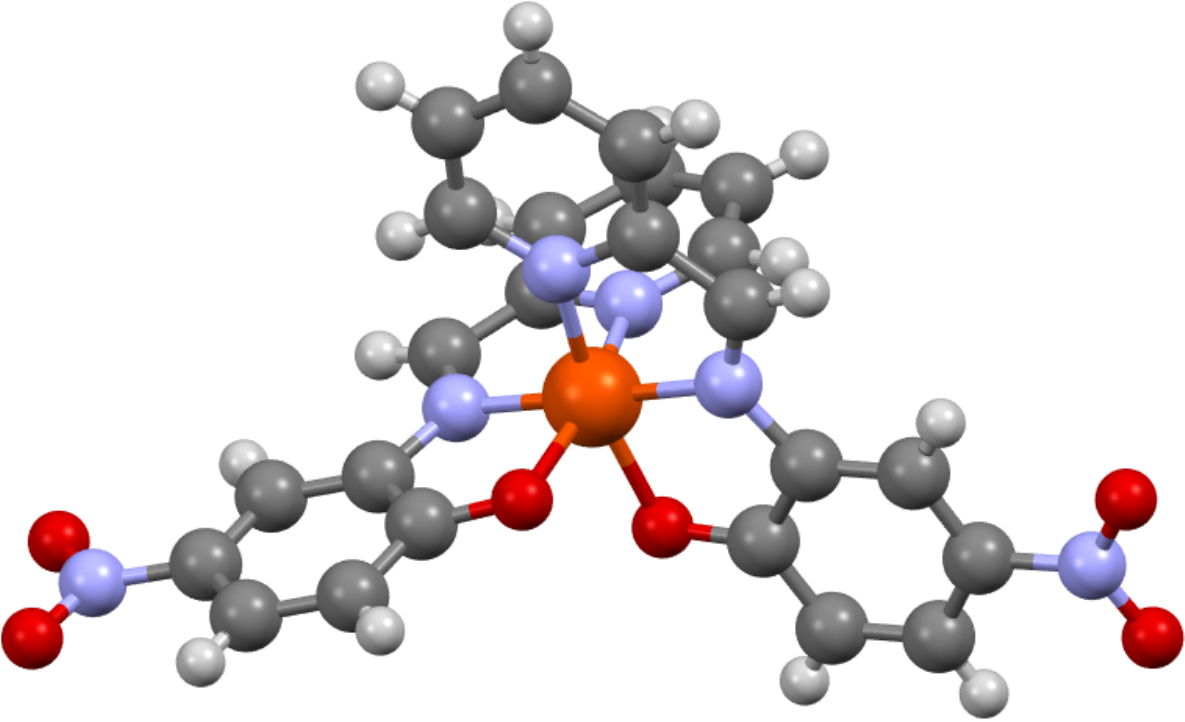}
    \caption{Molecular structure of Fe(pap-5NO$_2$)$_2$. Fe (orange), O (red), N (blue), C (grey), and H (white).}
    \label{fig:molecule}
\end{figure}

The compound [Fe(pap-5NO$_2$)$_2$] crystallizes in a monoclinic structure (space group C2/c), undergoing an isostructural spin-crossover transition that preserves the symmetry while inducing significant changes in the lattice parameters as shown in Table \ref{tab:crystal_data_pap}. Crystallographic studies \cite{Iasco2015-ov} show that it consists of a mononuclear Fe(II) complex coordinated by two tridentate Schiff-base ligands (Hpap-5NO$_2$), derived from the condensation of pyridine-2-carbaldehyde with 2-hydroxy-5-nitroaniline. Each ligand coordinates through N,N,O donor atoms, leading to an asymmetric octahedral coordination environment of type FeN$_4$O$_2$, as illustrated in Fig. \ref{fig:molecule}.

\begin{table}[ht]
\centering
\caption{Crystal data and structure refinement parameters for [Fe(pap-5NO$_2$)$_2$] (1) in the LS (100 K) and HS (320 K) states.}

\resizebox{\linewidth}{!}{%
\begin{tabular}{lcc}
\hline
\hline
 & 100 K (LS) & 320 K (HS) \\
\hline
Chemical formula & C$_{24}$H$_{16}$FeN$_6$O$_6$ & C$_{24}$H$_{16}$FeN$_6$O$_6$ \\
Formula weight (g/mol) & 540.28 & 540.28 \\
Crystal system & monoclinic & monoclinic \\
Space group & C2/c & C2/c \\
$a$ (\AA) & 22.948(2) & 23.267(18) \\
$b$ (\AA) & 9.0760(8) & 8.902(7) \\
$c$ (\AA) & 11.9420(9) & 11.967(9) \\
$\beta$ (°) & 119.587(2) & 114.198(13) \\
$V$ (\AA$^3$), $Z$ & 2162.9(3), 4 & 2261(3), 4 \\
Density (calc) (g cm$^{-3}$) & 1.659 & 1.587 \\
Absorption coefficient (mm$^{-1}$) & 0.756 & 0.670 \\
Goodness-of-fit on $F^2$ & 1.041 & 1.016 \\
$R_1$, $wR_2$ [$I > 2\sigma(I)$] & 0.0465, 0.0833 & 0.0431, 0.0992 \\
$R_1$, $wR_2$ (all data) & 0.0881, 0.0961 & 0.0718, 0.1130 \\
$\Delta\rho_{\mathrm{min}}$, $\Delta\rho_{\mathrm{max}}$ (e \AA$^{-3}$) & $-0.457$, $0.420$ & $-0.411$, $0.280$ \\
\hline
\hline
\end{tabular}%
}

\label{tab:crystal_data_pap}
\end{table}

The coordination sphere is composed of two nitrogen atoms from the imine groups, two nitrogen atoms from the pyridine rings, and two oxygen atoms from phenolate groups, forming a distorted octahedral geometry. Structural parameters obtained from single-crystal X-ray diffraction reveal typical spin-state-dependent bond lengths, with Fe–O, Fe–N$_{\mathrm{im}}$, and Fe–N$_{\mathrm{py}}$ distances increasing from approximately 1.98, 1.88, and 1.96 Å in the low-spin (LS) state to 2.07, 2.12, and 2.22 Å in the high-spin (HS) state, respectively, reflecting the population of antibonding orbitals upon the LS $\rightarrow$ HS transition. The unit-cell volume variation associated with the spin transition is $\Delta V/V \approx 4.5\%$, evidencing a strong coupling between the electronic and structural degrees of freedom.

In addition, the Fe(II) oxidation state is confirmed by Mössbauer spectroscopy, which clearly distinguishes the LS (S = 0) and HS (S = 2) states through their characteristic quadrupole doublets, ruling out the presence of Fe(III) species and confirming that the spin crossover is metal-centered. It is also shown by David Gracia \textit{et. al.}\cite{Gracia2025-pj} that the SCO transition occurs near room temperature and shifts linearly to higher temperatures with increasing pressure, accompanied by an increase in thermal hysteresis. At ambient pressure, the transition is associated with an isobaric volume change $\Delta V_p = 48 \pm 4$ Å$^3$ (for $Z = 4$), corresponding to approximately 2.3$\%$ of the unit-cell volume, whereas under isothermal conditions an average volume change $\Delta V_T = 75 \pm 10$ Å$^3$ was observed. These structural changes are directly linked to the barocaloric response, which reaches isothermal entropy changes of up to 70 J kg$^{-1}$ K$^{-1}$ and adiabatic temperature variations of 14 K for $\Delta p = 2.0$ kbar. 

\section{Effective Hamiltonian}
In order to describe the spin-crossover behavior, we follow the Hamiltonian proposed by Babilotte and Boukheddaden \cite{Babilotte2020-cq}:
\begin{equation} 
\begin{aligned} H &= \frac{K}{2} \sum_{ij} \left\{ x_{ij} - \left[a_{HL} + \frac{\delta a}{4}(\sigma_i + \sigma_j)\right]\right\}^2 \\ &\quad + P \sum_{ij} x_{ij} + \Delta_{eff} \sum_i \sigma_i \label{eq:hamiltonian}
\end{aligned}
\end{equation}
where the first and second terms account for lattice elasticity and the pressure effect, respectively. The third term describes the effective energy gap $\Delta_{eff}$ between the $e_g$ and $t_{2g}$ orbitals, where $\sigma = \pm 1$ represents the high-spin ($\sigma = +1$) and low-spin ($\sigma = -1$) states. Finally, $\delta a$ is the lattice parameter difference between the HS and LS states, whereas $a_{HL}$ corresponds to the metal–metal atomic distance.

As also discussed by Reis et al. \cite{Reis2024}, after assuming a homogeneous distance between the metal centers ($x_{ij} = x$) and applying mean-field theory at the equilibrium position, the Hamiltonian in Eq.~\ref{eq:hamiltonian} can be used to obtain the virtual magnetization ($m = \expval{\sigma}$) as a function of temperature ($T$) for a given pressure ($P$):
\noindent
\begin{equation}
m(T,P) = \tanh \left(\frac{m(T,P) \cdot \gamma}{t} + \ln \mathcal{G}\right)
\label{eq:virtual_mag}.
\end{equation}

In the above relation, $\gamma = (\Delta_0 + P \delta V / 4 k_B)/T_0$ and $t = T/T_0$, where $\Delta_0$, $T_0$, and $\mathcal{G}$ are fitting parameters, and $\delta V$ is the volume change between the low-spin and high-spin configurations. The parameter $\mathcal{G} = \mathcal{G}_{HS}/\mathcal{G}_{LS}$ represents the degeneracy ratio between high and low-spin states.

In this model, $\mathcal{G}$ is treated as a numerical fitting parameter that does not depend on temperature or pressure. However, {Ribeiro et al.}\cite{Ribeiro2019-zw} and {von Ranke et al.}\cite{Von_Ranke2022-fm} proposed an alternative approach in which this factor explicitly depends on spin and phonon contributions, therefore{,} {becoming} pressure and temperature dependent.

Now that the virtual magnetization has been defined, the high-spin{(low-spin)} molar fraction $n_{HS}$($n_{LS} = 1 - n_{HS}$) can be obtained. In this framework, if $m = +1${($-1$)}, all molecules are in the HS{(LS)} configuration. Therefore, the high-spin molar fraction can be written as:

\begin{equation}
n_{HS}(T,P) = \frac{m(T,P) + 1}{2}.
\end{equation}

As expected, this quantity varies between 0 and 1, corresponding to the cases where all molecules are in the LS or HS state, respectively.

\section{Barocaloric Effect}

Once the high-spin molar fraction is determined, the barocaloric effect can be evaluated. Here, we focus on the entropy description of the material, which consists of three {main} contributions:
\begin{equation}
S (T,B,P) = S_c(T, P) + S_s(T,B, P) + S_{o}(T, P)
\end{equation}

The terms {on the right} correspond to the configurational, spin{,} and lattice entropies, respectively. Since we are interested in the pressure range of a few {kilobar}, the acoustic contribution to the entropy is usually neglected. Therefore, if the material undergoes a change in pressure, the corresponding entropy variation can be written as:
\begin{equation}
\Delta S (T,B,\Delta P) = S (T,B, P_f)-S (T,B, P_i)
\end{equation}
where $P_f$ and $P_i$ correspond to the final and initial applied pressures, respectively.
\subsection{Configurational entropy}
The configurational entropy is related to the spatial distribution of high and low{-}spin molecules and is obtained by minimizing the free energy of the system with respect to temperature, yielding:
\begin{equation}
S_c(T,P) = -R \big[ n_{HS} \ln(n_{HS}) + (1 - n_{HS}) \ln(1 - n_{HS}) \big]
\label{eq:entropy_config}
\end{equation}

Usually, when pressure is applied, the spin crossover temperature ($T_{1/2}$) increases proportionally to the applied pressure. As a result, $n_{HS}(P_f) < n_{HS}(P_i)$ over the entire temperature range. Therefore, for a given $\Delta P$, the associated $\Delta S_c$ typically reduces (increases) the maximum entropy for $T > T_{1/2}$ ($T < T_{1/2}$).
\subsection{Spin entropy}
    {By its turn}, the spin entropy is given by the weighted sum of the spin entropies of high and low-spin molecules, normalized by their respective molar fractions. This contribution is obtained from the magnetic partition function and can be derived by minimizing the free energy with respect to the applied magnetic field, leading to:
\begin{align}
S_s(T,B,P) &= R\Bigg\{n_{HS}\Big[\ln (Z^s_{HS})-x_{HS}m_{HS}\Big] \\
    &\quad +(1-n_{HS})\Big[\ln (Z^s_{LS})-x_{LS}m_{LS}\Big]
\Bigg \},
\label{eq:entropy_spin}
\end{align}
with the partition function {$Z^s_{\alpha}$}:
\begin{equation}
    Z^s_\alpha(T,B)=
\frac{\sinh\left[\left(\frac{2J_\alpha+1}{2J_\alpha}\right)x_\alpha\right]}
{\sinh\left(\frac{x_\alpha}{2J_\alpha}\right)}.
\end{equation}
where $x_\alpha = g_\alpha J_\alpha \mu_B B / k_B T$, $J_\alpha$ is the total angular momentum and {$\alpha = \text{HS}, \text{LS}$}. Since we assume that the orbital angular momentum is quenched in both spin configurations, it is, $L_{HS} = L_{LS} = 0$, the Landé factor is $g_{HS} = g_{LS} = 2$. Moreover, because the low-spin state corresponds to $S_{LS} = 0$, its magnetic contribution vanishes. Therefore, the spin entropy variation is entirely due to the high-spin magnetization, weighted by its corresponding molar fraction.
\subsection{Lattice entropy}
The lattice entropy due to optical phonons is calculated considering that each mode $\nu$ behaves as a quantum harmonic oscillator with energy $\epsilon_\alpha = \sum_\nu (n_\nu + 1/2)\hbar \omega_\nu^\alpha$, where $\alpha$ stands for LS or HS state. Therefore, the entropy associated with molecular vibrations is:
\begin{equation}
S_o(T,P) = n_{HS} S^o_{HS}(T,P)+ (1 - n_{HS})S^o_{LS}(T,P)
\label{eq:entropy_lattice}
\end{equation}
where
\begin{equation}
S^{o}_{\alpha}(T,P) = R \sum_{\nu}
\left\{\frac{\beta\hbar \omega_{\nu}^{\alpha}}{e^{\beta \hbar \omega_{\nu}^{\alpha}} - 1} - \ln \left( 1 - e^{-\beta \hbar \omega_{\nu}^{\alpha}} \right)\right\}.
\label{eq:entropy_lattice_exp}
\end{equation}

Although Gábor Molnár \textit{et al.}\cite{molnar2019} showed that low-frequency modes ($< 400$ cm$^{-1}$), associated with the $3n - 6 = 15$ vibrational modes of an octahedron, contribute significantly to the lattice entropy, it is generally difficult to identify and isolate these 15 octahedral modes. Therefore, alternative approaches are often employed to address this issue. For instance, Reis and co-workers \cite{Reis2024} calculated the normal modes of the isolated molecule [FeL$_2$][BF$4$]$2$ using Gaussian 09 \cite{Gaussian09}{,} constructed a frequency histogram, in which a significant shift in the mean frequency was observed, from 264 cm$^{-1}$ in the LS state to 72 cm$^{-1}$ in the HS state, {and} for subsequent barocaloric analysis all 15 modes were assumed to be degenerate in both states. On the other hand, Sergi Vela \textit{et al.}\cite{Vela2025-yo} showed that, in order to accurately calculate the spin crossover temperature $T_{1/2}$, it is necessary to include not only the internal molecular modes but also the crystal vibrational modes, for which Quantum ESPRESSO \cite{Giannozzi2017-kr} was employed to capture the nonlinear behavior of $T_{1/2}$ as a function of pressure. In our approach, as discussed {with} more {details} in section {\ref{sec:comp}}, we calculated the normal modes of {an} isolated molecule, {containing} $n = 53$ atoms and therefore $3n - 6 = 153$ vibrational modes, which were subsequently used in the lattice entropy expression given {by} Eq. {(\ref{eq:entropy_lattice_exp})}.
\section{Computational details\label{sec:comp}}

All calculations were performed using the ORCA quantum chemistry package \cite{Neese2012-rd}. Geometry optimization and vibrational frequency calculations were carried out at the density functional theory (DFT) level employing the hybrid functional PBE0 \cite{Adamo1999-vt}, combined with Grimme’s D3 dispersion correction with Becke–Johnson damping (D3BJ) \cite{Grimme2011-vs}.

The electronic structure was described using the def2-TZVP \cite{Weigend2005-zg} basis set for all atoms, along with the corresponding auxiliary basis set def2/J \cite{Weigend2006-dc} to accelerate Coulomb integral evaluation within the resolution-of-identity (RI) approximation. Geometry optimizations and frequency calculations were performed with tight convergence criteria, allowing up to 500 iterations in both the self-consistent field (SCF) and geometry optimization procedures to ensure full convergence.

Vibrational frequencies were computed analytically at the optimized geometry which confirmed the nature of the stationary point as a true minimum (absence of imaginary frequencies). All calculations were performed in the gas phase considering an isolated molecule. A total charge of 0 was adopted in both cases. Spin multiplicity 5 was used to describe the high-spin (HS) state, while spin multiplicity 1 was employed for the low-spin (LS) state. The corresponding input files used in these calculations are provided in the Supporting Information.

To further investigate the correspondence between the high-spin and low-spin vibrational frequencies, we first computed the overlap between their respective displacement vectors, defined as:
\begin{equation}
    U_{ij} = \sum_{\alpha=1}^{N} 
    \mathbf{u}_{i,\alpha}^{\mathrm{HS}} 
    \cdot 
    \mathbf{u}_{j,\alpha}^{\mathrm{LS}},
    \label{eq:overlap}
\end{equation}
where $U_{ij}$ denotes the overlap between the i-th high-spin and j-th low-spin displacement modes, computed as a sum over atomic contributions ($\alpha$), with $\mathbf{u}_{i,\alpha}$ representing the three-dimensional displacement vector of atom $\alpha$. Since a given HS mode may exhibit significant overlap with more than one LS mode, a one-to-one correspondence between HS and LS frequencies is not always well defined. To account for this mode mixing, we associate each i-th HS mode with an effective LS frequency obtained as an overlap-weighted average of LS frequencies:
\begin{equation}
f^{\mathrm{LS}}_{\mathrm{eff},i} =
\frac{\sum_j w_{ij} f_j^{\mathrm{LS}}}
{\sum_j w_{ij}},
\label{eq:freq_eff}
\end{equation}
where $f_j^{\mathrm{LS}}$ is the frequency of the $j$-th LS mode and $w_{ij}=|U_{ij}|^2$ is the corresponding overlap weight.

\section{Results and Discussion}

The structural parameters extracted from both experimental and theoretical data provide a consistent and physically meaningful description of the spin-state dependence of the coordination environment in [Fe(pap-5NO$_2$)$_2$] as shown in Table \ref{tab:structural_parameters_exp_theo}. 
In the low-spin (LS) state, the average {iron-ligand (}Fe–L{)} bond distance is found to be 1.939 Å experimentally and 1.942 Å theoretically, indicating excellent agreement between the two approaches. This short bond length is characteristic of the LS configuration of Fe(II), where the $t_{2g}^{6}e_{g}^{0}$ electronic configuration leads to minimal antibonding interactions and, consequently, a more compact coordination sphere. In contrast, the high-spin (HS) state exhibits a substantial increase in the average Fe–L distance to 2.139 Å (Exp.) and 2.132 Å (Theo.), reflecting the population of antibonding $e_g$ orbitals in the $t_{2g}^{4}e_{g}^{2}$ configuration. This expansion of approximately 0.20 Å is fully consistent with the expected behavior for Fe(II) spin-crossover systems and is well reproduced by the theoretical model.

\begin{table}[htb!]
\centering
\caption{Selected Fe--L bond distances (\AA), cis L--Fe--L bond angles (°), average Fe--L distances, and angular distortion parameter $\Sigma$ for [Fe(pap-5NO$_2$)$_2$] in the LS and HS states.}
\begin{tabular}{lcccc}
\hline
\hline
 & \multicolumn{2}{c}{LS} & \multicolumn{2}{c}{HS} \\
\cline{2-3}\cline{4-5}
Parameter & Exp. (100 K) & Theo. & Exp. (320 K)& Theo. \\
\hline
Fe--O1 & 1.976 & 1.965 & 2.073 & 2.047 \\
Fe--N2 & 1.884 & 1.894 & 2.122 & 2.157 \\
Fe--N1 & 1.957 & 1.968 & 2.223 & 2.241 \\
O1--Fe--O1 & 87.63 & 90.63 & 97.69 & 100.92 \\
O1--Fe--N2 & 96.64 & 94.43 & 111.56 & 117.71 \\
O1--Fe--N2 & 83.97 & 83.91 & 77.17 & 77.06 \\
O1--Fe--N1 & 91.70 & 89.32 & 93.13 & 92.15 \\
N2--Fe--N1 & 81.52 & 81.77 & 74.05 & 73.43 \\
N2--Fe--N1 & 97.90 & 99.83 & 96.80 & 90.88 \\
N1--Fe--N1 & 92.62 & 94.28 & 89.95 & 89.08 \\
\hline
Fe--L$^{a}$ & 1.939 & 1.9423 & 2.139 & 2.132 \\
$\Sigma^{a}$ & 66.5 & 63.49 & 128.3 & 124.37 \\
\hline
\hline
\end{tabular}
\label{tab:structural_parameters_exp_theo}
\begin{flushleft}
$^{a}$ Fe--L is the average of the six Fe--ligand bond distances. 
$\Sigma = \sum_{i=1}^{12}|90-\phi_i|$, where $\phi_i$ are the twelve cis L--Fe--L angles.
\end{flushleft}
\end{table}

Beyond the isotropic expansion, the angular distortion parameter $\Sigma$ provides further insight into the geometric response of the coordination sphere. In the LS state, $\Sigma$ amounts to 66.5° experimentally and 63.49° theoretically, indicating a moderately distorted octahedral geometry. Upon transition to the HS state, $\Sigma$ increases dramatically to 128.3° (Exp.) and 124.37° (Theo.), revealing a pronounced enhancement of angular distortions. This significant increase demonstrates that the spin transition is accompanied not only by a uniform expansion of the Fe–L bonds but also by a substantial loss of octahedral symmetry.

The simultaneous increase in both Fe–L distances and $\Sigma$ highlights the strong coupling between electronic configuration and molecular geometry in this system. The occupation of antibonding orbitals in the HS state weakens the metal–ligand interactions in an anisotropic manner, allowing greater flexibility of the coordination sphere and facilitating deviations from ideal octahedral angles. This effect is particularly evident in the widening distribution of cis angles observed in the HS state, as reflected by the larger $\Sigma$ value.

Importantly, the close agreement between experimental and theoretical values for both Fe–L and $\Sigma$ confirms the reliability of the computational approach in capturing not only the average structural features but also the subtle angular distortions associated with the spin crossover and minor discrepancies can be attributed to thermal effects, crystal packing interactions, and the intrinsic limitations of the theoretical model, which considers an isolated molecule.
\begin{figure}[htb!]
    \centering
    \includegraphics[width=\linewidth]{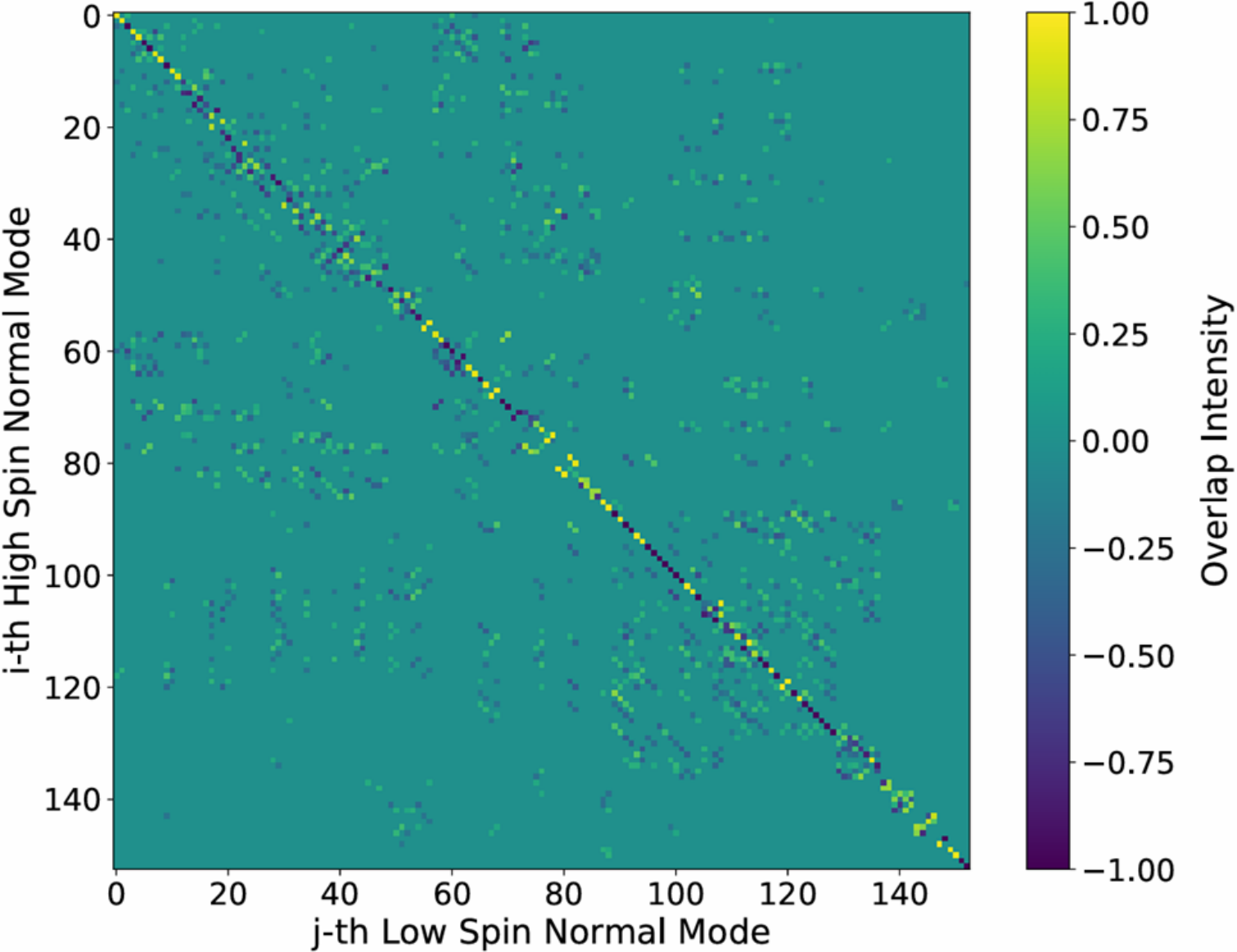}
    \caption{Overlap between the i-th high-spin normal mode and the j-th low-spin normal mode displacement vectors. Large values indicate a strong match between the normal modes, meaning that the vibrations occur in the same direction, whereas small values indicate that the normal modes are orthogonal and therefore unlikely to be matched.}
    \label{fig:overlap_matrix}
\end{figure}

\begin{figure*}[t]
    \centering
    \includegraphics[width=\textwidth]{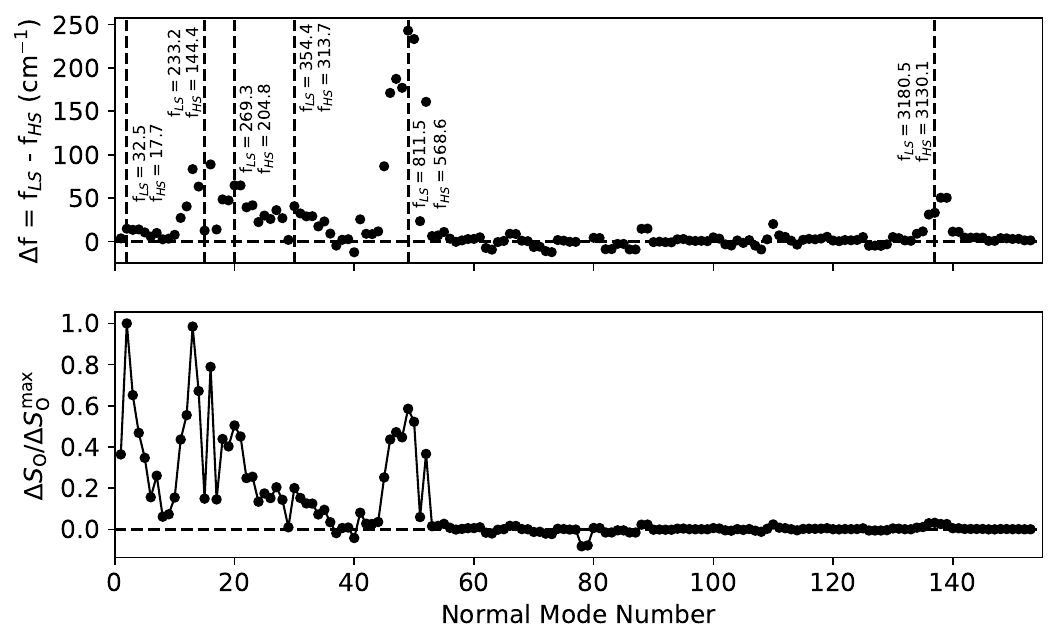}
    \caption{Top panel: Frequency differences between matched low-spin and high-spin normal modes as a function of the normal mode number. The vertical dashed lines indicate the peaks corresponding to the largest frequency differences within each region, along with their respective frequencies. Bottom panel: Normalized optical entropy difference, where values close to 0 indicate a smaller contribution to the total optical entropy, whereas values closer to 1 indicate a larger contribution.}
    \label{fig:freq_diff}
\end{figure*}

Based on Eq. \ref{eq:overlap}, we constructed the overlap matrix between the HS and LS vibrational modes, as shown in Fig. \ref{fig:overlap_matrix}. At first glance, the low-frequency region (i,j $\leq$ 12) and the high-frequency region (i,j $\geq$ 54) exhibit only a weak mixing between degenerate normal modes. In contrast, the intermediate frequency range (12 $<$ i,j $<$ 54) reveals a nontrivial matching between the normal modes, which is associated with the octahedra coupled to the organic ligands. Moreover, the overlap intensity in this intermediate region reflects the HS configuration, as it exhibits a highly distorted hexacoordinated environment compared to the LS configuration. Consequently, the overlap values in this range do not reach the higher intensities observed in the low/high frequency regions. 

Therefore, in order to identify the vibrational modes that provide the largest contributions to the barocaloric effect, we analyze the frequency difference between the HS modes and their corresponding effective LS counterparts as a function of the normal-mode index. Figure \ref{fig:freq_diff} presents the frequency shift associated with each matched mode pair calculated using Eq. \ref{eq:freq_eff}. The largest deviations are observed for modes within the intervals 10–30 and 40–60, which are predominantly associated with metal–ligand vibrational motions. Significant frequency shifts are also found for modes with larger normal-mode indices, corresponding mainly to intramolecular vibrations of the molecular framework. Nevertheless, metal–ligand vibrations dominate the vibrational entropy variation, particularly those associated with the first peak in Fig. \ref{fig:freq_diff}, because as it is shown in Eq. {(\ref{eq:entropy_lattice_exp})}, the contribution of a vibrational mode to the lattice entropy decreases exponentially with increasing frequency. Consequently, although appreciable frequency shifts are also observed for high-frequency intramolecular modes, their impact on the entropy variation is considerably smaller due to the exponential suppression of their entropic contribution.

\begin{figure}[!htb]
    \centering
    \includegraphics[width=1\linewidth]{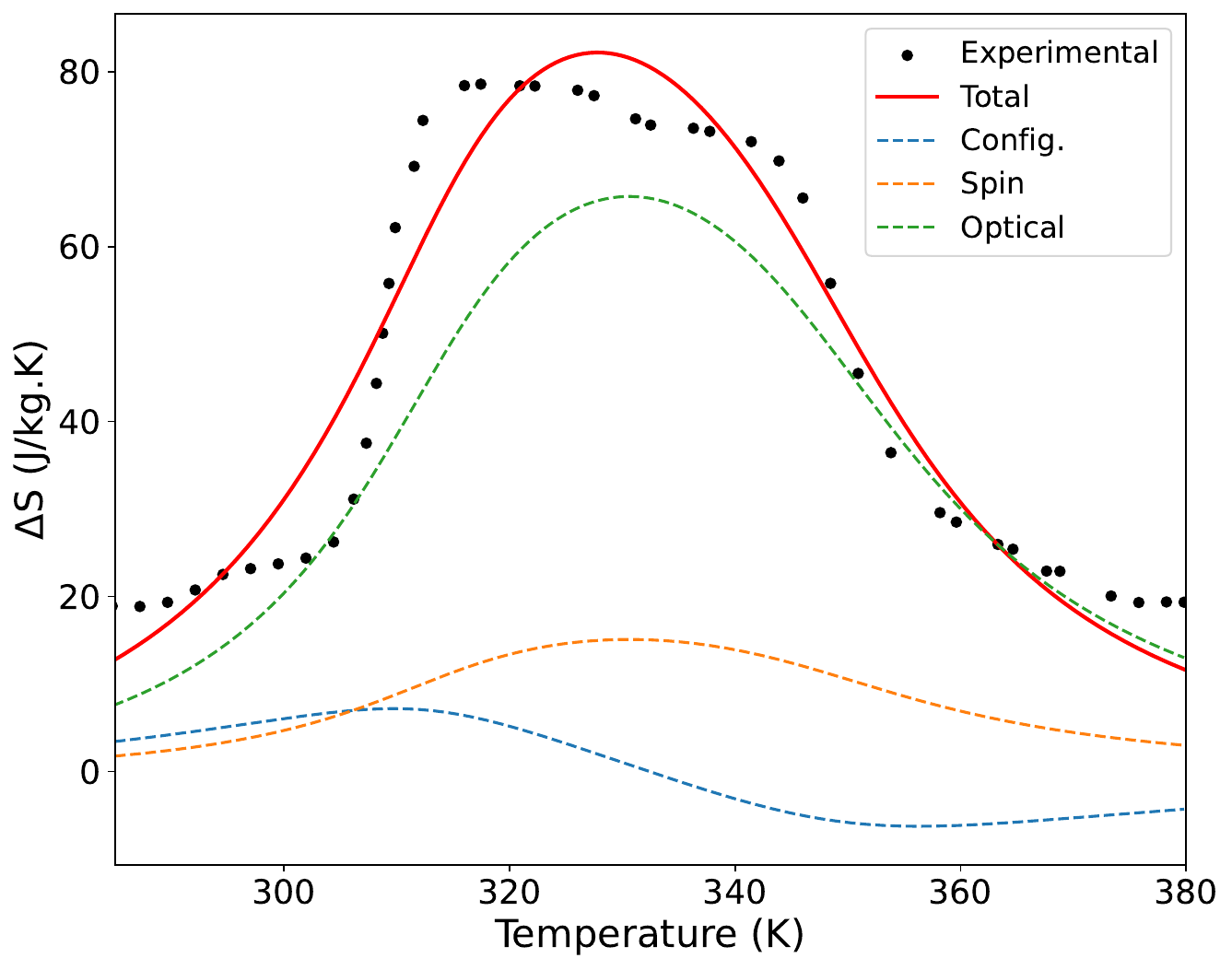}
    \caption{Experimental \cite{Gracia2025-pj} (solid circles) and calculated (solid lines) barocaloric entropy change. The total entropy variation is shown by the red solid line, whereas the blue, yellow, and green solid lines represent the configurational, spin, and optical contributions, respectively. The calculations were performed using fitted parameters $\mathcal{G}$ = 1457, $\Delta_0$ = 2285 K, and T$_0$ = 162 K.}
    \label{fig:ds_fit}
\end{figure}

The fit of the theoretical model to the experimental entropy-change data is presented in Fig. \ref{fig:ds_fit}, where the vibrational frequencies employed in the calculations were obtained from the DFT analysis discussed previously. A volume change $\delta V = 75$ Å$^3$ was considered for a pressure variation $\Delta p = 2$ kbar, as reported by Gracia \textit{et al.} \cite{Gracia2025-pj}. The set of parameters that yielded the best agreement with the experimental data consists of a degeneracy ratio $\mathcal{G} = 1457$, a characteristic temperature $T_0 = 162$ K, and a ligand-field energy $\Delta_0 = 2285$ K.

As shown in Fig. \ref{fig:ds_fit}, the mean-field model reproduces the experimental barocaloric entropy change remarkably well over the entire temperature range investigated. The decomposition of the total entropy variation reveals that the optical contribution constitutes the dominant component of the spin-crossover entropy change, accounting for approximately 84\% of the total entropy variation. This result is consistent with the substantial modifications in the vibrational spectrum induced by the spin-state transition and highlights the central role played by lattice dynamics in the barocaloric response of the compound. By its turn, the spin contribution is less pronounced, since it effectively arises only from the fraction of molecules populating the HS state. As for the configurational entropy, it exhibits a distinct behavior compared with the other contributions: below the transition temperature, the molecular population is predominantly in the LS state, whereas above the transition temperature, the HS state becomes predominantly populated.

Furthermore, by employing Eq. \ref{eq:virtual_mag} under ambient pressure and imposing the condition $m = 0$, together with the fitted parameters obtained from the entropy analysis, it is possible to estimate the theoretical spin-crossover transition temperature. The resulting value, $T_{1/2} = 313$ K, is in excellent agreement with the experimental transition temperature of 308 K.
\section{Conclusion and future perspectives}

In this work, we investigated the barocaloric effect in the spin-crossover complex Fe(pap-5NO$_2$)$_2$ by combining density functional theory calculations with an effective mean-field Hamiltonian. The optimized molecular structures reproduce the main experimental features of the low-spin and high-spin states, including the increase in the average Fe-L bond distance and the enhanced octahedral distortion associated with the spin transition.

The vibrational analysis revealed significant modifications in the normal-mode spectrum upon the SCO transition. By comparing the displacement-vector overlap between HS and LS modes, we showed that the strongest mode mixing occurs mainly in the intermediate-frequency region, where metal-ligand vibrations are coupled to motions of the organic ligands. The use of a singular value decomposition procedure allowed us to define effective LS frequencies associated with each HS mode and to identify the vibrational modes that contribute most strongly to the entropy variation.

The theoretical entropy-change curve obtained from the mean-field model is in very good agreement with the experimental barocaloric data for a pressure variation of $\Delta p = 2$ kbar. The best-fit parameters, $\mathcal{G} = 1457$, $\Delta_0 = 2285$ K, and $T_0 = 162$ K, also provide a theoretical transition temperature of $T_{1/2} = 313$ K at ambient pressure, in excellent agreement with the experimental value of 308 K. The decomposition of the total entropy change shows that the optical lattice contribution is dominant, accounting for approximately 84\% of the total barocaloric response, while configurational and spin contributions are comparatively smaller.

These results demonstrate that the large barocaloric effect observed in Fe(pap-5NO$_2$)$_2$ is mainly driven by vibrational entropy changes induced by the spin-crossover transition. In particular, the softening and redistribution of molecular vibrational modes associated with the metal–ligand vibrational motions between the LS and HS states play a central role in determining the pressure-induced entropy variation. Therefore, our results provide microscopic insight into the thermodynamic origin of the barocaloric response in this compound and reinforce the importance of accurately describing vibrational degrees of freedom in theoretical models of spin-crossover materials.

As future perspectives, the present approach could be extended by including crystal packing effects, pressure dependent phonon spectra, and low-frequency intermolecular modes, which are not fully captured in isolated-molecule calculations. Such improvements may provide an even more complete description of the barocaloric response and help guide the design of new spin crossover materials with enhanced solid-state refrigeration performance.


\begin{acknowledgements}

AS acknowledges financial support from CAPES (grant No. 88887.001022/2024-00) and from the CAPES Institutional Doctoral Sandwich Program Abroad (PDSE, Project No. 88881.220440/2025-01). MSR expresses gratitude to FAPERJ and CNPq for financial support. VS acknowledges financial support from FAPERJ (grants E-26/210.069/2023 and E-26/211.951/2021) and CNPq (grant 308200/2025-0). This research used resources of the National Energy Research Scientific Computing Center (NERSC), a Department of Energy User Facility (project m1503).

\end{acknowledgements}
\bibliographystyle{unsrt}  
\bibliography{bib}  

\end{document}